\documentclass[reprint,amsmath,amssymb,aps,twocolumn, superscriptaddress]{revtex4-2}
\usepackage{hyperref}
\usepackage{graphicx}
\usepackage{xcolor}
\usepackage{natbib}

\newcommand{\diff}{\ensuremath{\operatorname{d}\!}}
\def\HLTP{\hat{\mathcal{H}}^{\mathrm{LTP}}}
\def\ansp#1{\hat{a}_{\mathbf{#1}}}
\def\crsp#1{\hat{a}_{\mathbf{#1}}^{\dag}}
\def\anlo#1{\hat{b}_{\mathbf{#1}}}
\def\crlo#1{\hat{b}_{\mathbf{#1}}^{\dag}}

\def\anel#1{\hat{c}_{#1}}
\def\crel#1{\hat{c}_{#1}^{\dag}}

\newcommand{\freqn}[2]{\omega_{\mathbf{#2}}^{\mathrm{#1}}}
\newcommand{\freqp}[1]{\omega_{\mathbf{#1}}^{\pm}}
\newcommand{\rabi}[1]{\Omega_{\mathbf{#1}}}
\newcommand{\hopp}[1]{\mathrm{X}_{\mathbf{#1}}^{\pm}}

\begin{document}

\title{Non-equilibrium Electrical Generation of Surface Phonon Polaritons}

\author{Christopher R. Gubbin}
\affiliation{Sensorium Technological Laboratories, Nashville, TN, USA}
\author{Stanislas Angebault}
\affiliation{Sensorium Technological Laboratories, Nashville, TN, USA}
\author{Joshua D. Caldwell}
\affiliation{Sensorium Technological Laboratories, Nashville, TN, USA}
\affiliation{Vanderbilt University, Nashville, TN, USA}
\author{Simone De Liberato}
\affiliation{Sensorium Technological Laboratories, Nashville, TN, USA}
\date{\today}

\begin{abstract}
Notwithstanding its relevance to many applications in sensing, security, and communications, electrical
generation of narrow-band mid-infrared light remains highly challenging. Unlike in the ultraviolet or visible spectral regions {\textcolor{black}{few materials possess direct electronic transitions}} in the mid-infrared and most that do are created through complex band-engineering schemes.
An alternative mechanism, independent of dipole active material transitions, relies instead on energy lost to the polar lattice through the Coulomb interaction. Longitudinal phonons radiated in this way can be spectrally tuned through the engineering of polar nanostructures and coupled to localized optical modes in the material, allowing them to radiate mid-infrared photons into the far-field.
A recent theoretical work explored this process providing for the first time an indication of its technological relevance when compared to standard thermal emitters. In order to do so it nevertheless used an equilibrium model of the electron gas, making this model difficult to inform the design of an optimal device to experimentally observe the effect.
The present paper removes this limitation, describing the electron gas using a rigorous, self-consistent, non-equilibrium Green's function model, accounting for variations in material properties across the device, and electron-electron interactions. Although the instability of the Schrodinger-Poisson iteration limits our studies to the low-bias regime, our results demonstrate emission rates comparable to that of room-temperature thermal emission despite such low biases. 
These results provide a pathway to design a confirmatory experiment of this new emission channel, that could power a new generation of mid-infrared optoelectronic devices.

\end{abstract}

\maketitle

\section{Introduction}
\label{sec:Introduction}
As a consequence of strong, long-range Coulombic interactions between charged electrons and the crystal lattice, electrical transport in polar materials is dominated by longitudinal optical (LO) phonon scattering, which ultimately determines the electron saturation velocity \cite{ridleyHotphononinducedVelocitySaturation2004, khurginAmplifiedSpontaneousEmission2016a}. 
Recent works have proposed that LO phonons generated in this way can be put to use: in nanoscopic layers they are able to couple directly to light through shared boundary conditions \cite{gubbinOpticalNonlocalityPolar2020a,gubbinImpactPhononNonlocality2020a,gubbinNonlocalScatteringMatrix2020a,gubbinPolaritonicQuantizationNonlocal2022a,gubbinQuantumTheoryLongitudinalTransverse2022a}, leading to the far-field emission of mid-infrared photons \cite{ gubbinElectricalGenerationSurface2023}.\\
Polar dielectric crystals support localized excitations termed surface phonon polaritons (SPhPs). These modes are a consequence of strong coupling between photons and the polar optical phonon modes of the lattice \cite{caldwellLowlossInfraredTerahertz2015, hillenbrandPhononenhancedLightMatter2002}, existing in the Reststrahlen region separating the lattice longitudinal and transverse optical phonon frequencies in which the dielectric function turns negative. It is within this spectral range that the real part of the dielectric permittivity tensor becomes negative. As hybrid photon-phonon quasi-particles, they enable deep sub-diffractional confinement of energy \cite{sumikuraHighlyConfinedSwitchable2019, woessnerHighlyConfinedLowloss2015} and fine spectral tunablity \cite{luEngineeringSpectralSpatial2021, wangOpticalPropertiesSingle2013, dubrovkinResonantNanostructuresHighly2020,gubbinTheoreticalInvestigationPhonon2017, ellisAspectratioDrivenEvolution2016, gubbinStrongCoherentCoupling2016a} with applications in nonlinear optics \cite{gubbinTheoryNonlinearPolaritonics2017, gubbinTheoryFourWaveMixingPhonon2018,kitadeNonlinearShiftPhononPolariton2021, razdolskiSecondHarmonicGeneration2018a}, sensing \cite{autoreBoronNitrideNanoresonators2018, berteSubnanometerThinOxide2018} and near-field imaging \cite{kiesslingSurfacePhononPolariton2019, taubnerNearfieldMicroscopySiC2006}. The phonon-like nature of SPhPs has wide consequences for mid-infrared optoelectronics \cite{gubbinSurfacePhononPolaritons2022}, particularly in the design of tailored narrowband thermal emitters enabled by the mode's underlying spectral tunablity and narrow linewidth \cite{greffetCoherentEmissionLight2002, schullerOpticalAntennaThermal2009, arnoldCoherentThermalInfrared2012a}. \\
Longitudinal and transverse optical phonons are also tunable. In macroscopic systems the dispersion of each is determined by the composition and symmetries of the crystal lattice, but when the lengthscale approaches the phonon propagation length they can become localized. In this regime the continuous phonon dispersion evolves into a discrete one, analogous to the shift in dispersion experienced by photons in a Fabry-P\'erot resonator. The SPhPs in such a system exhibit spectral shifts \cite{gubbinOpticalNonlocalityPolar2020a} and the emergence of new modes observable from the far-field \cite{gubbinHybridLongitudinaltransversePhonon2019}. In this regime they acquire a new moniker: longitudinal-transverse polaritons (LTPs) \cite{gubbinOpticalNonlocalityPolar2020a, gubbinNonlocalScatteringMatrix2020a, gubbinImpactPhononNonlocality2020a, gubbinQuantumTheoryLongitudinalTransverse2022a} reflecting the mixture of longitudinal and transverse degrees of freedom and the consequent necessity of considering the true microscopic nature of the underlying phonons.\\ 
Longitudinal-transverse polaritons are able to couple directly to electrons through the Coulomb interaction and radiate into the far-field like SPhPs. This enables them to act as the active mode in an electrically driven emitter without the need for dipole-active electronic transitions. Optoelectronic devices designed in this way could provide a simple and flexible pathway to generating radiation: polar dielectric materials have a wide range of phonon frequencies spanning the mid-infrared spectral region and for each material LTP modes can be fine-tuned across the Reststrahlen spectral window. 
Electrical emission via LTPs is thus a complex multi-particle process, in which non-equilibrium electrons couple to LO phonons, which in turn couple to SPhPs, which themselves couple to free-space radiation.
{\textcolor{black}{To understand the practicality of LTP-based electrical emitters the previous study of LTP emission in polar nanosystems treated the electronic sub-system using a simple, equilibrium free-electron model characterized by a uniform electron temperature \cite{gubbinElectricalGenerationSurface2023}. Although such a model was sufficient to provide an estimation of the underlying efficiency of the LTP emission channel, it cannot account for local variations of the electron density, the effect of a multilayer structure with varying effective mass and conduction band offsets, or the effect of electron-electron repulsion. These shortcomings prevent calculation of the electrical current across the device, making it difficult to design and optimize a device to observe such an effect. \\
This Paper aims to overcome these issues, assessing LTP generation rates using self-consistently calculated electron distributions, determined by iterative solution of the coupled Schrodinger and Poisson equations. Our results allow us to assess the changes induced through a rigorous model of the electron gas. Note that the inherent instability in solving the Schrodinger-Poisson equation system \cite{zhu2024} limits this work to small bias, a regime where emission is weak. This regime is most interesting for an initial experiment as a sample exhibits less resistive heating, and resultant spurious thermal emission at small bias. Our results can be roughly extrapolated to larger bias, provides strong evidence for the feasibility of LTP based optoelectronic devices, and a pathway toward design of future experiments.}}

\section{Theoretical overview}

\subsection{Longitudinal-Transverse Polaritons in Thin Films}
\label{sec:LTPs}
Surface phonon polaritons are hybrid excitations, resulting from the strong coupling of optical phonons to light at the surface of a polar dielectric. Both components exhibit dispersion, meaning their frequency is a function of momentum, however the lengthscales for this dispersion are very different. A photon appreciably shifts frequency for wavevector changes comparable to it's inverse wavelength, while similar shifts for a phonon happen for wavevectors comparable to the inverse lattice spacing. This broad disparity often allows the optical response of an SPhP system to be accurately reconstructed neglecting the phonon dispersion. This is a local-response approximation, termed because the material response at a point $\mathbf{r}$ depends only on the local electric field at $\mathbf{r}$: a consequence of neglecting phonon dispersion is that all phonons are assumed to have zero momentum, so energy is solely transported in the photon field. In systems with length-scales approaching the phonon propagation length a local-response approximation can fail. Energy can be transported through optical phonon excitations, and the material response at $\mathbf{r}$ becomes nonlocal, depending on the driving field in a finite region enclosing $\mathbf{r}$ \cite{gubbinOpticalNonlocalityPolar2020a}. In this regime the hybrid nature of the mode becomes increasingly important. While in the local-response approximation SPhPs can be considered pure transverse excitations, with polarization fields $\mathbf{P}$ orthogonal to wavevector $\mathbf{Q}$  $(\mathbf{Q} \times \mathbf{P} = 0)$, in the nonlocal regime this is no longer the case leading to the existence of LTP excitations \cite{gubbinImpactPhononNonlocality2020a, gubbinNonlocalScatteringMatrix2020a, gubbinQuantumTheoryLongitudinalTransverse2022a}.\\
\begin{figure}
    \begin{center}
\includegraphics[width=0.4\textwidth]{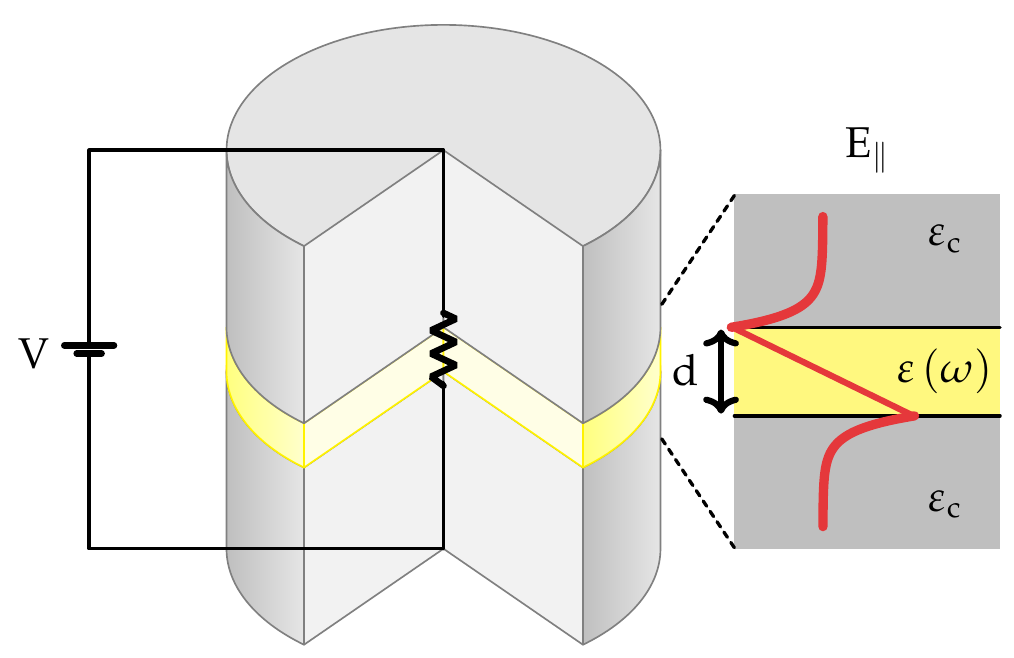}
    \end{center}
    \caption{Sketch of the photonic system under study. A polar dielectric film of thickness $d$ is sandwiched between symmetric high-index cladding layers characterized by non-dispersive dielectric constant $\varepsilon_\mathrm{c}$. In the Reststrahlen spectral region of the polar layer the system supports surface phonon polariton excitations, the mode profile for the epsilon-near-zero excitation is sketched in red. The device can emit light by applying a voltage perpendicular to the polar film as indicated.}\label{fig:Figure1}
\end{figure}\\
In this Paper we focus on a  nanoscale polar film of thickness $d$, sandwiched between non-polar cladding layers with a non-dispersive dielectric constant $\epsilon_\mathrm{c} > 0$ illustrated in Fig.\ref{fig:Figure1}. It has previously been demonstrated that LTPs in this system can be accurately characterized using a Hopfield model whose bare modes are the SPhP in the local-response approximation and the discrete LO phonon modes of the layer \cite{gubbinQuantumTheoryLongitudinalTransverse2022a}. The polar layer is characterized by an isotropic dielectric function
\begin{equation}
\epsilon\left(\omega\right) = \epsilon_\infty \frac{\omega_\mathrm{L}^2 - \omega\left(\omega + i \gamma\right)}{\omega_\mathrm{T}^2 - \omega\left(\omega + i \gamma\right)} \label{eq:localDielectricFunction},
\end{equation}
where $\epsilon_\infty$ is the high-frequency dielectric constant, $\omega_\mathrm{T} \; (\omega_\mathrm{L})$ is the zone-center transverse (longitudinal) optical phonon frequency and $\gamma$ is the damping constant, assumed here to be independent from frequency and polarization. \\
In the local-response approximation the film supports long-range and short-range SPhPs as a consequence of hybridization between the charge oscillations at each of its interfaces. The mode profile of the long-range epsilon-near-zero (ENZ) mode is sketched in Fig.~\ref{fig:Figure1}. In the thin-film limit the ENZ mode dispersion as a function of the in-plane wavevector $\mathbf{q}$ is
\begin{equation}
    \freqn{SP}{\mathbf{q}} = \sqrt{\frac{\omega_\mathrm{L}^2 + \lvert \mathbf{q} \rvert x \omega_\mathrm{T}^2}{ 1 + \lvert \mathbf{q} \rvert x}}, \label{eq:SPhPDispersion}
\end{equation}
where $x = d \epsilon_\mathrm{c} / 2 \epsilon_\infty$ \cite{gubbinQuantumTheoryLongitudinalTransverse2022a}, while localized LO phonons supported by the film satisfy the quadratic approximation \cite{gubbinPolaritonicQuantizationNonlocal2022a}
\begin{equation}
    \omega_{\mathbf{q}, n}^{\mathrm{L}} = \sqrt{\omega_\mathrm{L}^2 - \beta_\mathrm{L}^2 \left(\lvert \mathbf{q}\rvert^2 + \zeta_n^2\right)}, \label{eq:localizedDispersion}
\end{equation}
in which $\beta_\mathrm{L}$ is the phonon velocity in the limit of quadratic dispersion, and $\zeta_n = \frac{2 (n - 1) \pi}{d}, \; (n \in \mathbb{Z}^+)$ is the quantized out-of-plane wavevector of the localized mode. For the remainder of this paper we focus on the case where only one $n=1$ localized phonon is near resonance with the ENZ, dropping the index from $\freqn{\mathrm{L}}{\mathbf{q}}$ and $\zeta$. The LTP spectrum in this system is well characterized using a two-oscillator Hopfield model, given in the rotating-wave approximation by
\begin{equation}
	\HLTP = \hbar\sum_{\mathbf{q} } \biggr \{   \freqn{SP}{\mathbf{q}} \crsp{q} \ansp{q} 
	 +  \freqn{L}{q} \crlo{ q} \anlo{q} 
         + \rabi{q} \left(\ansp{q}\crlo{q}+\crsp{q}\anlo{q}  \right)\biggr\}, \label{eq:RWAHamiltonian}
\end{equation}
where $\crsp{q}$ and $ \crlo{q}$ are bosonic creation operators for the SPhP and the $n=1$ localized LO phonon mode, and the Rabi frequency is given by \cite{gubbinQuantumTheoryLongitudinalTransverse2022a}
\begin{equation}
    \lvert \rabi{\mathbf{q}} \rvert^2 =  \frac{2 \beta_{\mathrm{L}}^2}{d^2} 
        \frac{
            (\omega_{\mathrm{L}} - \freqn{SP}{q})
            (\omega_{\mathrm{L}} + \freqn{SP}{q})
            }{
                \freqn{SP}{q} \freqn{L}{q}
            }.\label{eq:RabiFrequency}
\end{equation}
The frequencies of the two resulting LTP branches, which we will refer to as lower (-) and upper (+) polaritons, can be found by calculating the 
eigenvalues of the Hamiltonian in Eq.~\ref{eq:RWAHamiltonian}
\begin{align}
    \freqp{q} =& \biggr[\frac{\freqn{SP \; 2}{\mathbf{q}} + \freqn{L \; 2}{\mathbf{q}}}{2} \nonumber \\
               & \pm \frac{1}{2}\sqrt{\left[\freqn{SP \; 2}{\mathbf{q}} - \freqn{L \; 2}{\mathbf{q}} \right]^2 + 4 \Omega_{\mathbf{q}}^2 \freqn{SP}{\mathbf{q}} \freqn{L}{\mathbf{q}}}\biggr]^{1/2}.
    \label{eq:polfreq}
\end{align}
Polaritons can be understood considering the eigenvectors of the Hamiltonian, which describe the mixture of underlying modes in the coupled excitation. The fraction of each bare mode in a given polariton is termed the Hopfield coefficient, which is given by the square of the relevant element of the eigenvector. The Hopfield coefficient describing the longitudinal content of each polariton branch can be written as
\begin{equation}
    \lvert \hopp{q} \rvert^2 = \frac{1}{2} \left[ 1 \pm \frac{\Delta \omega_{\mathbf{q}}}{\sqrt{\Delta \omega_{\mathbf{q}}^2 + 4 \rabi{q}^2}}\right], \label{eq:Hopfield}
\end{equation}
while the one describing the transverse content is $\lvert \mathrm{Y}_\mathbf{q}^\pm \rvert^2 = 1 - \lvert \hopp{q} \rvert^2$. Here $\Delta \omega_{\mathbf{q}} = \freqn{SP}{q} - \freqn{L}{q}$ is the frequency difference between the bare mode frequencies.

\subsection{Non equilibrium Green's functions}
\label{sec:GF}
Transport in open systems can be described using a Landauer-B\"{u}ttiker formalism \cite{buttikerFourTerminalPhaseCoherentConductance1986}. In an open-dissipative quantum system electrons transition between energy and momentum states by interacting with external reservoirs, requiring knowledge of two-time correlation functions for the electron gas. Correlation functions can be calculated in a non-equilibrium Green's function (NEGF) formalism \cite{schwingerBrownianMotionQuantum1961, keldyshDiagramTechniqueNonequilibrium2023, caroliDirectCalculationTunneling1971, meirLandauerFormulaCurrent1992, dattaNanoscaleDeviceModeling2000}. Green's functions generalize the time-dependent electron distribution $n\left(\mathbf{k}, t\right)$ to a two-time correlation function between
creation and annihilation operators $\crel{\mathbf{k}}\left(t\right)$ and $ \anel{\mathbf{k}}\left(t\right)$ for an electron with wavevector $\mathbf{k}$ at time $t$.  The NEGF formalism describes the electronic population utilizing the distributional lesser and greater Green's functions \cite{lakeSingleMultibandModeling1997}
\begin{subequations}
\begin{align}
	G^<\left(\mathbf{k, k}'; t, t'\right) & = i \langle c_{\mathbf{k}'}^\dag \left(t'\right), c_{\mathbf{k}}\left(t\right)\rangle, \label{eq:Lesser}  \\
	G^>\left(\mathbf{k, k}'; t, t'\right) & = - i\langle c_{\mathbf{k}} \left(t\right), c_{\mathbf{k}'}^\dag \left(t'\right)\rangle, \label{Greater}
\end{align}
\end{subequations}
which in the steady-state regime considered in this Paper solely depend on the time difference $\tau = t - t'$.
Fourier transforming with respect to space and time we obtain the
 distributional Green's functions 
 $G^{\lessgtr}\left(\mathbf{r, r}; E\right)$ describing correlations between the amplitudes at positions $\mathbf{r, r'}$ with energy $E$, 
 whose diagonal element yields the real-space carrier density in the device
\begin{equation}
    n\left(\mathbf{r}\right) = - \frac{2 i}{\Delta \mathrm{A}} \int \frac{1}{2\pi} G^{<}\left(\mathbf{r, r}; E\right) \mathrm{d E}, \label{eq:Charge}
\end{equation}
where we wrote the element volume as the product of quantization area $\mathrm{A}$ and mesh-spacing $\Delta$ \cite{lakeSingleMultibandModeling1997}. \\
Calculating $G^\lessgtr$ requires knowledge of the associated dynamic Green's functions: the 
retarded Green's function $G^R$ and the advanced Green's function $G^A = \left[G^R\right]^{\dag}$. These can be derived from a generalization of the Schrodinger equation to an open system, contacted to external reservoirs using leads characterized by self-energies quantifying the complex frequency shift associated with the coupling 
\begin{equation}
	\left[E  - \mathcal{H}_0 - \Sigma^R  \right] G^{R} = \delta\left(\mathbf{r - r'}\right), \label{eq:Retarded}
\end{equation}
where $\mathcal{H}_0$ is the closed system Hamiltonian discussed in Appendix~\ref{appendixsec:Hamiltonian} and $\Sigma^R$ is the retarded self-energy whose derivation for coupling to semi-infinite leads is described in Appendix.~\ref{appendixsec:Leads}. The distributional Green's functions $G^\lessgtr$ are then given by the kinetic equation
\begin{align}
	G^{\lessgtr} & = G^R \Sigma^{\lessgtr} G^A, \label{eq:Kinetic}
\end{align}
where $\Sigma^\lessgtr$ are lesser and greater self-energies. Self-energies refer to inflows and outflows of charge from the distributions described by $G^\lessgtr$ and their difference relates to the level broadening $\Gamma$
\begin{equation}
    \Gamma = i \left[\Sigma^R - \Sigma^A\right] = i\left[\Sigma^> - \Sigma^<\right]. \label{eq:Broadening}
\end{equation}
Note that as $\mathcal{H}_0$ is dependent on the electrostatic potential $\phi$ through the Poisson equation, whose solution is outlined in Appendix.~\ref{appendixsec:Potential}, which in turn depends on the carrier density calculated through Eq.~\ref{eq:Charge}, an iterative solution is necessary, as outlined in Appendix.~\ref{appendixsec:Loop}.\\

\subsection{Figures of Merit}
\label{sec:FiguresOfMerit}
In this Paper we aim to compare the efficiency of the longitudinal phonon driven electrical LTP emission channel with those of SPhP thermal emitters that are already well established in the literature \cite{greffetCoherentEmissionLight2002, arnoldCoherentThermalInfrared2012a, luEngineeringSpectralSpatial2021}. For this reason we follow the discussion in Ref.\cite{gubbinElectricalGenerationSurface2023}, defining two figures of merit for the performance of an electrical emitter that rely on comparison of electrical and thermal populations at the same wavevector, rather than net emitted fluxes. The first is the ratio between the steady-state electrically excited LTP population $N_\mathbf{q}^{\mathrm{ss}}$ in a lattice at ambient temperature $T_a = 300$ K and thermal population $N_0$ {\textcolor{black}{at the same temperature
\begin{equation}
    F^{(1)} \left(\omega_\mathbf{q}, T_a, T_e \right) = \frac{N^{\mathrm{ss}}\left(\omega_\mathbf{q}, T_a\right)}{N^0\left(\omega_\mathbf{q}, T_a\right)}. \label{eq:fom1}
\end{equation}
This represents the emission enhancement compared to the thermal case, which may be observed in an experiment. The thermal population $N_0$ is given by the Bose-Einstein distribution
\begin{equation}
    N^0\left(\omega, T\right) = \frac{1}{e^{\hbar \omega / k_\mathrm{B} T} - 1}.
    \label{eq:BEDist}
\end{equation}}
The steady-state population of excitations in the LTP reservoir under interaction with the electron gas derived in Ref.~\cite{gubbinElectricalGenerationSurface2023} is found using the Fermi golden rule to be
\begin{equation}
N^\mathrm{ss}\left(\omega_\mathbf{q}, T\right) = \frac{N^0\left(\omega_\mathbf{q}, T\right) \gamma_\mathbf{q} + \Gamma_{\mathbf{q}}^{\mathrm{in}}}{\gamma_{\mathbf{q}} + \Gamma_{\mathbf{q}}^\mathrm{out} - \Gamma_\mathbf{q}^\mathrm{in}}. \label{eq:SteadyStatePopulation}
\end{equation}
where $T$ is the temperature, $\omega_\mathbf{q}$ is the reservoir excitation frequency at in-plane wavevector $\mathbf{q}$, $\gamma_\mathbf{q}$ is the wavevector-dependent damping rate for excitations in the reservoir, and $\Gamma_\mathbf{q}^{\mathrm{in} / \mathrm{out}}$ are the rates of in/out-scattering from the \emph{reservoir} due to interactions with the electron gas. Note that Eq.~\ref{eq:SteadyStatePopulation} is equally valid for excitation of localized LO phonons and LTPs provided the appropriate dispersion relation and electron-phonon interaction rates.\\
Although the figure of merit is the same as that considered in Ref.~\cite{gubbinElectricalGenerationSurface2023}, the key difference is that in this work the electron distributions entering $\Gamma^{\mathrm{in/out}}$ are calculated self-consistently using a NEGF formalism. Additionally, compared to the free-electron model used in Ref.~\cite{gubbinElectricalGenerationSurface2023} our theory requires a new parameter. In a free-electron model the carrier distribution is characterized by a single temperature that defines both the shape of the Fermi surface and the drift velocity. In this work the two properties are decoupled: the electron gas exists at ambient temperature regardless of the bias across the device. This improves the accuracy of our results. We take the electronic temperature to be ambient $T_a = 300$ K in all results in this Paper.\\
The second figure of merit quantifies the number of photons generated by a single electron passing through the system. The photon-fraction of an LTP can be approximated as ratio of it's group-velocity to the speed of light
\begin{equation}
    \xi_\mathbf{q}^{\pm} = \frac{1}{c} \biggr \| \frac{\mathrm{d} \omega_\mathbf{q}^\mathrm{SP}}{\mathrm{d} \mathbf{q}} \biggr \| \lvert \mathrm{Y}_\mathbf{q}^\pm \rvert^2,
\end{equation}
where $\mathrm{Y}$ is the transverse mode Hopfield coefficient.
Modes with flat dispersion are matter-like, with low group velocity and vanishing $\xi_\mathbf{q}$, while fast dispersing excitations have a sizeable photonic component and acquire a finite $\xi_\mathbf{q}$. {\textcolor{black}{The electrically generated \emph{photonic} power per unit surface is then given by
\begin{equation}
    F^{(2)} = \frac{1}{2 \pi} \int \mathrm{d}q\, q\, \hbar \omega_{\mathbf{q}} \xi_\mathbf{q} \Gamma_\mathbf{q}^\mathrm{net}, \label{eq:fom2}
\end{equation}
where $\Gamma_{\mathbf{q}}^{\mathrm{net}} = \left[1 + N\left(\omega_\mathbf{q}, T_l\right)\right]\Gamma_\mathbf{q}^{\mathrm{in}} - N\left(\omega_\mathbf{q}, T_l\right) \Gamma_\mathbf{q}^\mathrm{out}$ is the net excitation rate of LTPs.}

\begin{figure}
    \begin{center}
        \includegraphics[width=0.5\textwidth]{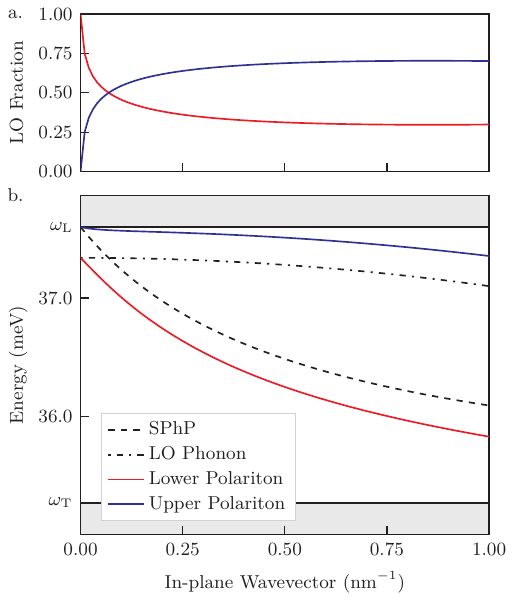}
    \end{center}
    \caption{The two-oscillator  dispersion of longitudinal-transverse polaritons for a $3$ nm AlGaAs film. a. Longitudinal Hopfield coefficient in each branch from Eq.~\ref{eq:Hopfield} b. Dash-dotted (dashed) line indicates the dispersion of the $n=1$ localized phonon from Eq.~\ref{eq:localizedDispersion} (SPhP from Eq.~\ref{eq:SPhPDispersion}). Red and blue lines indicate LTP frequencies from Eq.~\ref{eq:polfreq}.  }\label{fig:Figure2}
\end{figure}

\begin{figure}
    \begin{center}
\includegraphics[width=0.5\textwidth]{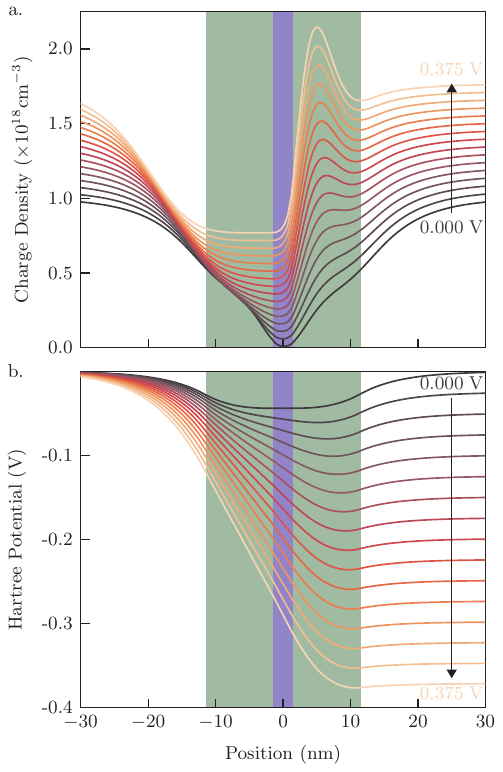}
    \end{center}
    \caption{Characteristics of a 3 nm AlGaAs barrier, sandwiched between GaAs leads as the applied bias is altered. The darkest curve represents the zero bias result, while lighter colors illustrate larger applied bias. The background shading illustrates different material regions of the device: blue indicates the 3 nm AlGaAs barrier, green the 10 nm lightly doped GaAs access regions and whithe the strongly doped GaAs leads. Panels show a. The self-consistent charge distribution. For clarity results for non-zero bias are offset vertically. b. The self-consistent Hartree potential.}\label{fig:Figure3}
\end{figure}

\begin{figure}
    \begin{center}
\includegraphics[width=0.5\textwidth]{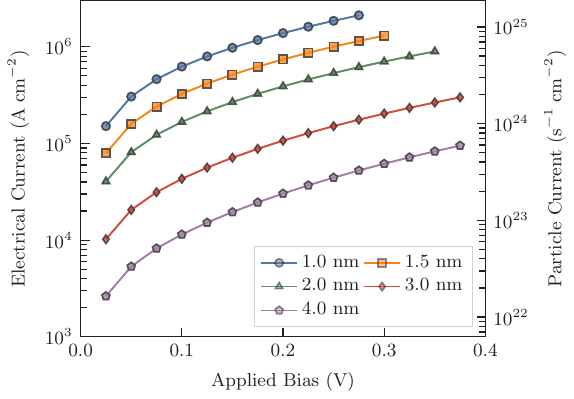}
    \end{center}
    \caption{Comparison of ballistic charge current as a function of barrier thickness. The left axis shows the electrical current density $J$ from Eq.~\ref{eq:BallisticCurrent} while the right shows the particle current density $J / e$.}\label{fig:Figure4}
\end{figure}

\section{Results}

\subsection{Electrical characterization}
\label{sec:System}
In this Paper we consider a system consisting of an $\mathrm{Al_{0.24}Ga_{0.76}As}$ barrier sandwiched between GaAs leads. The polar AlGaAs barrier is a negative dielectric near to it's LO phonon frequency $\omega_\mathrm{L}=37.6 \; \mathrm{meV}$ where it supports an ENZ SPhP excitation \cite{lugliInteractionElectronsInterface1992}. For simplicity we consider the GaAs access regions to be positive dielectrics characterized by high-frequency dielectric function $\epsilon_\infty = 10.89$. This approximation allows use of the analytical LTP theory presented in Section.~\ref{sec:LTPs}. Dispersion in the true dielectric function of the GaAs will shift the dispersion of the SPhP mode reducing the bandwidth, which can be accounted for in a full numerical calculation of the nonlocal optical response \cite{gubbinNonlocalScatteringMatrix2020a}. \\
In Fig.\ref{fig:Figure2}a we plot the localized phonon, or longitudinal content of the two LTP branches calculated through Eq.~\ref{eq:Hopfield}. At zero-wavevector the lower polariton is entirely longitudinal in nature, while the upper is pure transverse. At non-zero wavevector both branches can be considered mixed excitations. This can be understood considering the modal dispersion in Fig.~\ref{fig:Figure2}b calculated through Eq.~\ref{eq:polfreq}. At zero wavevector, polariton frequencies (solid lines) coalesce with those of the bare modes (black lines): the upper-polariton is entirely SPhP-like, while the lower is entirely LO-phonon-like. This is a consequence of the coupling in Eq.~\ref{eq:RabiFrequency} tending to zero as the SPhP mode frequency $\freqn{SP}{q} \to \omega_\mathrm{L}$. At larger wavevectors both modes acquire substantial longitudinal character \cite{gubbinQuantumTheoryLongitudinalTransverse2022a} as they deviate from the bare mode frequencies.\\
The photonic system is contacted by sandwiching between highly n-doped (doping density $N_{\mathrm{D}}=10^{18}\; \mathrm{cm}^{-3}$) semi-infinite GaAs leads. GaAs in the internal photonic system is considered to be lightly doped ($N_{\mathrm{D}}=10^{12}\;\mathrm{cm}^{-3}$). A buffer region thickness of $10$ nm is used in this Paper, sufficiently thick that the highly confined LTP does not extend into the strongly doped leads. These lightly doped buffer regions ensure the cladding dielectric function $\epsilon_\mathrm{c}$ remains positive, reduce maximal charge density in the central region, and reduces charge accumulation at the emitter. This improves current stability \cite{doTransportNoiseResonant2006} and convergence of the self-consistent calculation. The structure is characterized by effective mass $m^* = 0.063 \; (0.083) m_e$ with $m_e$ the bare electron mass, and static dielectric constant $\epsilon_{\mathrm{st}} = 12.90 \; (12.21)$ in the GaAs (AlGaAs) region. The system can be seen in the colors of Fig.~\ref{fig:Figure3}: blue indicates the AlGaAs film, green the lightly doped buffer region and white the heavily doped leads.\\
The conduction band offset of the AlGaAs barrier ($0.3\;\mathrm{eV}$) expels charge from the device central region. This can be observed in Fig.~\ref{fig:Figure3}a, which shows self-consistent calculations of the coherent charge distributions in the device for a $3$ nm barrier at ambient temperature ($T_a = 300$ K) using Eq.~\ref{eq:Charge}. At zero bias there is a deep minimum in the charge density at the center of the AlGaAs barrier. As the bias ramps up, electrons are able to tunnel through the barrier: this increases the charge density in the barrier somewhat but predominantly leads to a strong increase in charge density at the barrier edge as electrons with insufficient energy to tunnel across are pushed up against it. The corresponding Hartree potential in each case is shown in Fig.~\ref{fig:Figure3}b. In the leads the gradient of the potential tends to zero and the charge density approaches the donor density, while in the device center, where the potential gradient is large there is a strong deviation from the background charge distribution.  \\
{\textcolor{black}{Ballistic current conduction across the barrier is given in terms of the dynamic Green's functions and self-energies of the device contacts as
\begin{multline}
    J = \frac{2 e}{\hbar \mathrm{A}} \sum_\mathbf{k} \int \frac{\mathrm{d E}}{2 \pi} \mathrm{Tr}\left\{\Gamma^e G^R \Gamma^c G^A \right\} \\ \times \left[f_e\left(T_a, E\right) - f_c\left(T_a, E\right)\right], \label{eq:BallisticCurrent}
\end{multline}
where $f_{e} \; \left(f_c\right)$ is the Fermi function in the emitter (collector) contact and $\Gamma^{e} \; \left(\Gamma^c\right)$ is the broadening associated with the contact, calculated from the contact self-energies derived in Appendix.~\ref{appendixsec:Leads} through Eq.~\ref{eq:Broadening}. This is plotted for a selection of barrier thicknesses in Fig.~\ref{fig:Figure4}. As the applied bias is ramped up, inter-terminal current increases rapidly as more electrons acquire sufficient energy to tunnel across the barrier. Thinner barriers exhibit higher current at equal applied bias as the evanescent tails of wavefunctions belonging to electrons with energies below the barrier height are more able to penetrate through. On the large bias side of Fig.~\ref{fig:Figure4} the maximum voltage presented decreases with film thickness, this is a consequence of instability in the Schrodinger-Poisson iterations \cite{zhu2024}. Instability precludes us from finding a stable solution in reasonable time when the device is approaches an open circuit, a condition reached for smaller bias in thinner films.}}

\begin{figure*} 
    \begin{center}
 \includegraphics[width=0.85\textwidth]{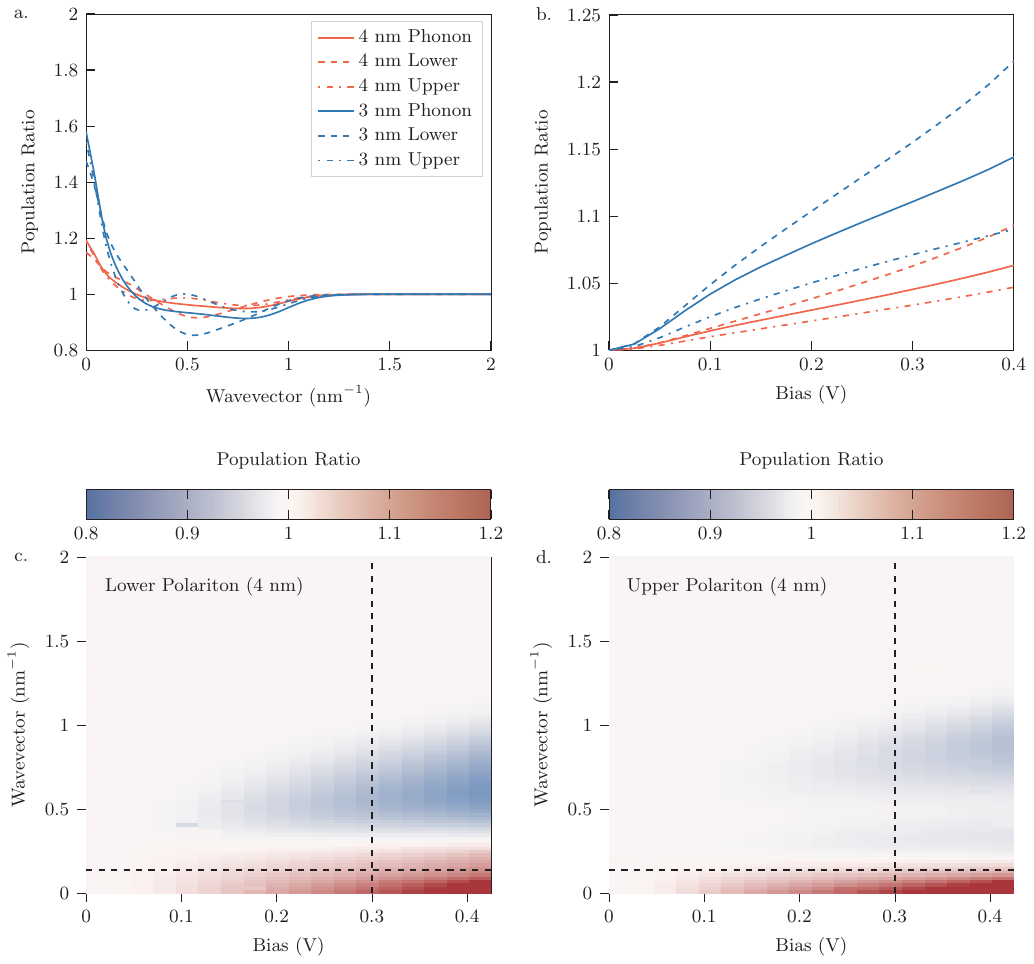}
    \end{center}
    \caption{Figure of Merit $F^{(1)}$ from Eq.~\ref{eq:fom1} evaluated for film thickness 3 nm (red) and 4 nm (green). Results for localized phonons in the absence of SPhP coupling are shown with solid lines, while those for the two solutions to Eq.~\ref{eq:RWAHamiltonian} are shown with dashed (lower polariton) and dash-dotted (upper polariton) lines respectively. Data is presented a. As a function of wavevector for fixed bias $0.3$ V. b. As a function of bias for fixed wavevector $0.02 \; \mathrm{nm}^{-1}$ c. As a colormap of both variables for the 4 nm sample's lower polariton branch. d. As a colormap of both variables for the 4 nm sample's upper polariton branch. Dashed lines in c. and d. indicate the cuts plotted in a. and b.}
    \label{fig:Figure5}
\end{figure*}
\begin{figure*}
    \begin{center}
        \includegraphics[width=0.85\textwidth]{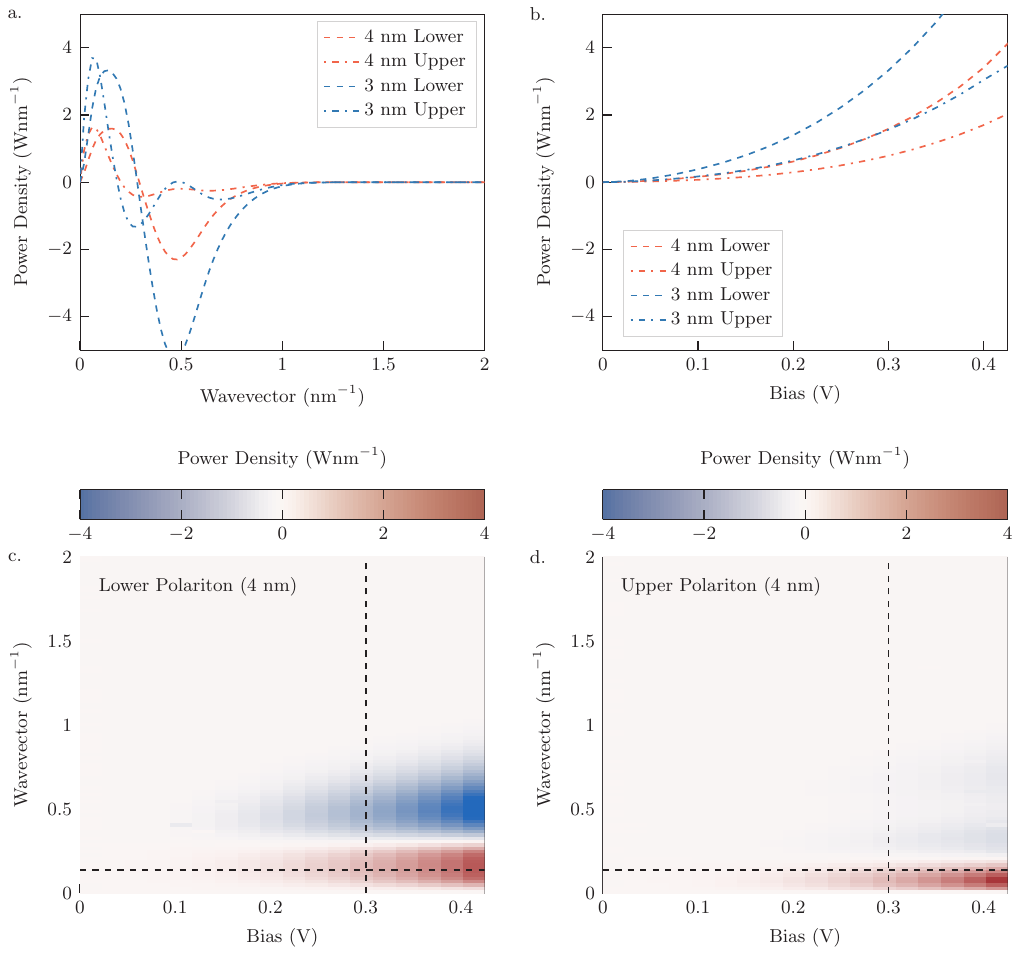}
    \end{center}
    \caption{Integrand for Figure of Merit $F^{(2)}$ from Eq.~\ref{eq:fom2} evaluated for film thicknesses 3 nm (red) and 4 nm (green). Results for localized phonons in the absence of SPhP coupling are shown with solid lines, while those for the two solutions to Eq.~\ref{eq:RWAHamiltonian} are shown with dashed (lower polariton) and dash-dotted (upper polariton) lines respectively. Data is presented as a. a function of wavevector for fixed bias $0.3$ V. b. As a function of bias for fixed wavevector $0.02 \; \mathrm{nm}^{-1}$. c. As a colormap of both variables for the 4 nm sample's lower polariton branch. d. As a colormap of both variables for the 4 nm sample's upper polariton branch. Dashed lines in c. and d. indicate the cuts plotted in a. and b.}
    \label{fig:Figure6}
\end{figure*}

\begin{figure}
    \begin{center}
        \includegraphics[width=0.45\textwidth]{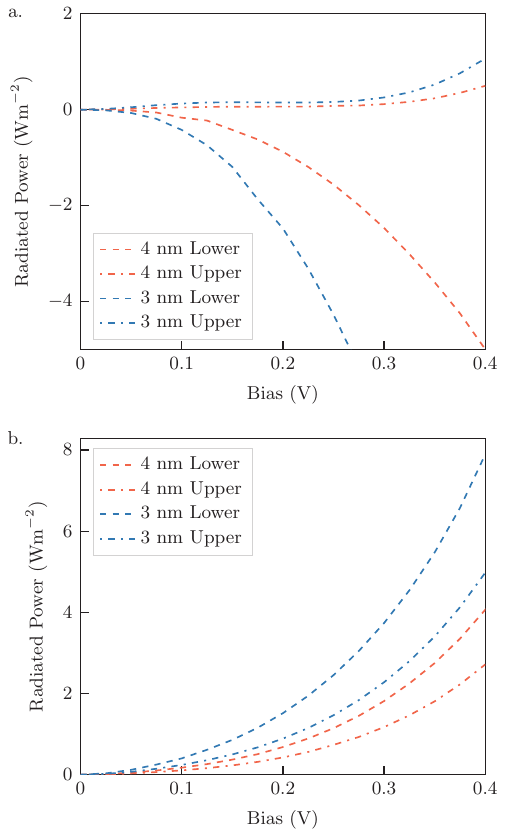}
    \end{center}
    \caption{Figure of merit $F^{(2)}$ integrated over the in-plane wavevector of the outgoing LTP for a. All wavevectors. b. Wavevectors less than $0.25\;\mathrm{nm}^{-1}$.}
    \label{fig:Figure7}
\end{figure}
\subsection{LTP emission}

In this section we consider incoherent transport in the device. We do not calculate a full-self consistent solution of both LTP and electron populations in the device, instead considering the perturbative regime where the equilibrium electron distribution radiates LTPs but is not appreciably altered by doing so. Firstly we formulate some useful figures of merit for assessing incoherent transport in the NEGF formalism. The total rate of out-scattering experienced by a state in the electron gas with in-plane momentum $\mathbf{k}$ and energy $E$ is proportional to the distributional greater Green's function
\begin{align}
    \Gamma_{i} \left(\mathbf{k}, E\right) &= \frac{1}{2 \pi \hbar} G_{ii}^>\left(\mathbf{k}, E\right) \Sigma_{ii}^<\left(\mathbf{k}, E\right),
    \label{eq:outscatter}
\end{align}
where quantities indexed $\mathbf{k}$ are those defined in Section.~\ref{sec:GF} at $\mathbf{k}' = \mathbf{k}$. For the device under consideration in Fig.~\ref{fig:Figure1}, the in-plane dimension is translationally invariant, which allows us to pass to Fourier space in the in-plane wavevector. The out-of-plane growth axis cannot be transformed in this way due to the spatial inhomogeneity introduced by the multilayer structure. Along this out-of-plane dimension quantities are defined on a discrete spatial mesh, rather than in the continuous coordinate system used in Section.~\ref{sec:GF}. Subscript indices $i$ refer to elements in the corresponding real-space matrix or vector, so $G_{ii}^>$ refers to the $i$th element of the greater Green's function diagonal, and $\Gamma_i$ refers to the $i$th element of the vector $\Gamma$. 
Note that in Eq.~\ref{eq:outscatter} we make a diagonal self-energy approximation $\Sigma_{ij}^< = 0 \; (i \neq j)$. Insodoing we implicitly assume the scattering process is local, this means our results give a conservative lower bound on the true scattering rate in an LTP based emitter. In a full model, electrons would be able to interact with phonons in the barrier while outside it \cite{kubisSelfconsistentQuantumTransport2007, ridleyElectronphononInteractionQuasitwodimensional1982, ridleyElectronScatteringConfined1989}.\\
The out-scattering rate has contributions from both emission and absorption of longitudinal excitations through self-energy $\Sigma^<$. The diagonal component of the lesser self-energy can be written in general form as
\begin{multline}
\Sigma_{ii}^<\left(\mathbf{k}, E\right) = \sum_{\mathbf{q}} \mathcal{K}_{i}\left(\mathbf{q}\right) \\
	\biggr\{N^0\left(\omega_\mathbf{q}, T_a
        \right) G_{ii}^<\left(\mathbf{k - q}, E - \hbar \omega_\mathbf{q}\right) \\
	+ \left[N^0\left(\omega_\mathbf{q}, T_a
\right) + 1\right] G_{ii}^<\left(\mathbf{k - q}, E + \hbar \omega_\mathbf{q}\right)\biggr\},
\end{multline}
where $\omega_{\mathbf{q}}$ is the frequency of the reservoir mode with in-plane wavevector $\mathbf{q}$, $\mathcal{K}$ is a process-specific scattering kernel and $i$ labels the discrete spatial indices in the mesh. The equilibrium Bose-Einstein population is given by $N^0$. Calculation of the self-energies for localized phonons is discussed in Appendix.~\ref{appendix:B}. \\
To compute the figures of merit defined in Section.~\ref{sec:FiguresOfMerit} it is necessary to derive the rates of in-scattering and out-scattering from the reservoir, which can be written taking appropriate components of $\Sigma^<$ and summing over the electronic degrees of freedom
\begin{subequations}
\begin{align}
    \Gamma_{\mathbf{q}}^{\mathrm{in}} &= \sum_i \mathcal{K}_i\left(\mathbf{q}\right)  \sum_{\mathbf{k}, E} G_{ii}^>\left(\mathbf{k}, E\right) G_{ii}^{<} \left(\mathbf{k - q}, E - \hbar \omega_{\mathbf{q}}\right),  \\
    \Gamma_{\mathbf{q}}^{\mathrm{out}} &= \sum_i \mathcal{K}_i\left(\mathbf{q}\right)  \sum_{\mathbf{k}, E} G_{ii}^> \left(\mathbf{k}, E\right) G_{ii}^{<} \left(\mathbf{k - q}, E + \hbar \omega_{\mathbf{q}}\right), 
    \label{eq:InOut}
\end{align}
\end{subequations}
where terms labeled ``$\mathrm{in}$'' refer to processes that increase the number of excitations in the reservoir, and those labeled ``$\mathrm{out}$'' refer to those that reduce the number of excitations in the reservoir. \\
As LTPs couple to electrons through their longitudinal phonon component, the theory used to calculate the localized phonon scattering rates in Appendix~\ref{appendix:B} can be extended to describe LTP scattering by substituting the localized phonon energies with eigenfrequencies from Eq.~\ref{eq:polfreq}, and weighting the contribution from each polariton branch by the Hopfield coefficient describing it's fractional longitudinal content Eq.~\ref{eq:Hopfield}
\begin{multline}
    \Sigma_{ii}^{< \; \pm} \left(\mathbf{k}, E\right) = \sum_{\mathbf{q}} \lvert U_{\mathbf{q}} \rvert^2 \lvert u(z_i) \rvert^2 \lvert \hopp{q} \rvert^2 \\
    \biggr\{ N\left(\freqp{q}, T_a\right) G_{ii}^< \left(\mathbf{k - q}, E - \hbar \freqp{q} \right)\\
        \left[N\left(\freqp{q}, T_a\right) + 1\right] G_{ii}^<\left(\mathbf{k-q}, E + \hbar \freqp{q} \right)
	\biggr\},
\end{multline}
where $u$ is the out-of-plane LO phonon wavefunction and $U_{\mathbf{q}}$ is the matrix element for the scattering process, both defined in Appendix~\ref{appendix:B}. The longitudinal phonon Hopfield coefficient $\hopp{q}$, defined in Eq.~\ref{eq:Hopfield} weights the sum by the longitudinal content of the mode, reducing the self-energy in regions of the dispersion where the LTP is photon-like. We can identify the scattering kernel for the LTP as
\begin{equation}
    \mathcal{K}_{i}^\pm \left(\mathbf{q}\right) = \lvert U_{\mathbf{q}} \rvert^2 \lvert u(z_i) \rvert^2 \lvert \hopp{q} \rvert^2,
\end{equation}
which is just that for the localized phonon in Eq.~\ref{eq:LOKernel} weighted by the squared longitudinal Hopfield coefficient.

{\textcolor{black}{We compute the first figure of merit $F^{(1)}$ using Eq.~\ref{eq:fom1}. The steady-state population in the denominator is taken to be that of the device operating under ambient conditions at $300\;\mathrm{K}$. Results are shown as a function of LTP wavevector for fixed applied bias $0.3$ V in Figure.\ref{fig:Figure5}a for two barrier thicknesses $3$ nm and $4$ nm. Solid lines show the figure of merit in the absence of SPhPs. In this case the scattering kernel is given by Eq.~\ref{eq:LOKernel} and the mode frequency is $\freqn{L}{q}$. The strongest enhancements are observed for wavevectors less than $0.25 \; \mathrm{nm}^{-1}$. In this region the figure of merit converges for all branches as the scattering rate significantly exceeds the damping rate. The dominant scattering channel varies as the wavevector increases, a result of the unique modal dispersion Fig.~\ref{fig:Figure2} as otherwise reduced phonon content in the polaritonic case would result in a lower scattering rate. The peak at small momentum is a significant new result and suggests that outcoupling of these excitations may be possible using grating features in excess of $100$nm. Note however, that the LTPs \emph{only} exist at significantly larger wavevectors than SPhPs and that the LO-SPhP coupling breaks down too close to zero wavevector. All wavevectors in Fig.~\ref{fig:Figure5} are large compared to the free-photon at this wavelength ($\approx 1 \; \mu \mathrm{m}^{-1}$). Thinner films exhibit nonlocality closer to the lightline, and will be most promising for optoelectronic applications \cite{gubbinPolaritonicQuantizationNonlocal2022a}, allowing LTPs to be outcoupled to the far-field with larger grating features.\\
The same figure of merit at fixed in-plane wavevector $0.14 \; \mathrm{nm}^{-1}$ is shown in Fig.~\ref{fig:Figure5}b The quantity increases in all cases as a function of applied bias at this wavevector. At larger bias more electrons are in states that are able to penetrate the barrier (Figure.~\ref{fig:Figure4}), leading to an increase in the steady-state LTP population. Compared to previous calculations of this quantity \cite{gubbinElectricalGenerationSurface2023} we predict weaker growth in $F^{(1)}$ as the bias increases. This is a consequence of the simple free-electron model presented previously, in which larger bias led to a linear increase in electron tunneling through the barrier, and a nearly linear increase in scattering. In this work while increasing bias does increase the current (Fig.~\ref{fig:Figure4}) most electrons still do not occupy states capable of tunneling through the barrier, rendering the figure of merit less sensitive to changes in the applied bias. An electron with $0.4$ V energy has an effective temperature $\approx 4000$ K, so despite the low bias considered here we nonetheless extend beyond the parameter range considered in \cite{gubbinElectricalGenerationSurface2023}. This reduction could be countered by considering a quantum well active region, or a device such as a resonant tunneling diode where phonon-electron overlap could be greatly enhanced \cite{doTransportNoiseResonant2006}. Note that at zero bias the scattering process deactivates and the thermal population is returned. At this point the device is thermalized. The electronic population continues to increase with bias, and in an experiment could be more readily observed by application of large bias in excess of that explored in this work. We limit our studies to biases below $0.45$ V because of the instability of the Schrodinger-Poisson outer iteration at large bias \cite{zhu2024} in the open circuit limit.\\
In Fig.~\ref{fig:Figure5}c and d we plot $F^{(1)}$ as a function of both wavevector and bias for the lower and upper polariton branches of the 4 nm film respectively. In both cases the strongest emission occurs close to zero wavevector, increasing monotonously with the applied bias. Emission at larger wavevectors is suppressed as the bias increases, at these wavevectors charge carriers absorb phonons more quickly than they emit reducing the steady-state population. Interaction with the electron gas shifts excitations in the reservoir from large wavevector states to ones closer to the light-line.}}

{\textcolor{black}{The second figure of merit describes the photonic power spectral density generated within the device. Figure.~\ref{fig:Figure6}a shows the integrand
\begin{equation}
    I_\mathbf{q} = \frac{\lvert \mathbf{q} \rvert}{2 \pi} \hbar \omega_\mathbf{q} \xi_\mathbf{q} \Gamma_\mathbf{q}^\mathrm{net},
\end{equation}
which describes a power density $\mathrm{W nm}^{-1}$. Results for the two film thicknesses as a function of excitation wavevector for fixed bias $0.3$ V are shown in Fig.~\ref{fig:Figure6}. Contrasting Fig.~\ref{fig:Figure5} the pure phonon result is omitted as it has no photonic component. At zero wavevector power emitted into both branches tends to zero, which occurs because the power density is linearly proportional to the modal wavevector. In Figure.~\ref{fig:Figure6}b we plot the same quantity as a function of bias at small wavevector $0.14\;\mathrm{nm}^{-1}$. At zero bias the system is thermalized and there is no net-emission. When the bias is increased, emission grows as more electrons penetrate the barrier with sufficient energy to emit LTPs. Emission is stronger into the lower polariton branch at this wavevector, a consequence of it's larger photonic component (Fig.~\ref{fig:Figure2}).\\
In the lower panels of Fig.~\ref{fig:Figure6} c and d we plot $I_\mathbf{q}$ as a function of both bias and wavevector for the lower and upper branch in the $4 \; \mathrm{nm}$ film respectively. As the bias increases low-wavevector photon emission increases, particularly for the upper polariton branch. As for $F^{(1)}$, the integrand $I_\mathbf{q}$ becomes negative at larger wavevector as the electron gas absorbs phonons from the reservoir.}}

{\textcolor{black}{Finally we plot the integrated photonic power density $F^{(2)}$. Results are shown in Fig.~\ref{fig:Figure7}a. The resulting value is positive for the upper-polariton branch for both thicknesses, but negative for the lower-polariton branch. This can be understood referring to Fig.~\ref{fig:Figure6}c-d. At larger wavevectors the spectral density becomes negative in both polariton branches, but this is enhanced in the lower branch. When integrating over the in-plane wavevector of the phonon this large wavevector absorption is enhanced compared to the small-wavevector emission. Taken over all wavevectors phonon absorption reduces the total photonic power generated compared to the thermal case.\\
Fortunately, a real device will not outcouple LTPs from every in-plane wavevector. By restricting the outcoupling mechanism to in-plane wavevectors less than $0.25 \; \mathrm{nm}^{-1}$ the photonic power density emitted from the device can be greatly increased as shown in Fig.~\ref{fig:Figure7}b. Longitudinal-transverse polaritons generated in this emission region could be addressed using a grating or similar.}}

\section{Conclusion}
Generation of narrowband mid-infrared light is an active problem in mid-infrared optoelectronics \cite{gubbinSurfacePhononPolaritons2022}. This Paper presented a novel method of designing narrowband mid-infrared electrical emitters without the need for complex band-engineering. Harnessing the natural Coulomb interaction between electrons and the optical phonon modes of polar crystal lattices we demonstrated the generation of photons, capable of coupling to the far-field through intermediary longitudinal-transverse polaritons (LTPs). Our results confirm previous works proposing LTP systems as electrically driven mid-infrared emitters \cite{gubbinElectricalGenerationSurface2023}, extending to account for the electronic structure of the device in addition to the LTP dynamics. Although we only study small applied bias in this work, we show significant changes in the LTP population which could be further increased in a device operating under larger voltage. Our theory can be enhanced using established techniques to calculate the charge distribution in the presence of scattering in the self-consistent Born approximation \cite{lakeSingleMultibandModeling1997}, and to calculate gain in such a system \cite{wackerGainQuantumCascade2002, leeNonequilibriumGreenFunction2002}. Our initial results can be built on to design experimental LTP-based emitters, and the presented figures of merit can be used to assess their likely performance.  They underline that this new channel of emission could underpin a new generation of mid-infrared optoelectronic emitters powered by longitudinal-transverse polaritons.

\section{Acknowledgements}
J. D. C. would like to acknowledge support from ONR Spector: Office of Naval Research Grant N00014-22-12035.

\newpage
\appendix
\numberwithin{equation}{section}
\renewcommand{\theequation}{\thesection\arabic{equation}}
\section{Self-consistent Charge Calculation}
\label{appendix:A}
This Appendix summarizes the ingredients necessary to calculate the self-consistent electronic charge distribution in the device in the absence of scattering between energy and momentum states within the device.

\subsection{Calculation of the Closed System Hamiltonian}
\label{appendixsec:Hamiltonian}
The dynamic retarded Green's functions are calculated using Eq.~\ref{eq:Retarded}. This equation requires prior knowledge of the Hamiltonian of the closed electronic system $\mathcal{H}_0$ and the self-energies associated with coupling to the external leads of the device $\Sigma^R$.\\
We consider transport in a system homogeneous in the $xy$ plane, but inhomogeneous along the transport direction $z$. The closed system, defined in the absence of coupling to the leads, is described by one-dimensional Hamiltonian
\begin{equation}
	\mathcal{H}_0 = - \frac{\hbar^2}{2} \left[\frac{\diff}{\diff z} \frac{1}{m^*\left(z\right)} \frac{\diff}{\diff z}\right] + V_{k_\parallel}\left(z\right) + \frac{\hbar^2 k_\parallel^2}{2 m_\mathrm{L}^*},
\end{equation}
where $m^*(z)$ is the effective mass at position $z$, $m_\mathrm{L}^*$ is the effective mass in the semi-infinite left-lead, $k_\parallel$ is the in-plane wavevector and
\begin{equation}
	V_{k_\parallel}(z) = V\left(z\right) + \frac{\hbar^2 k_\parallel^2}{2 m_\mathrm{L}^*} \left[\frac{m_\mathrm{L}^*}{m^*\left(z\right)} - 1 \right],
\end{equation}
where $V(z)$ is the on-site potential due to conduction band offsets and the Hartree potential. \\
We discretize the Hamiltonian in the nearest-neighbour tight-binding approximation, whose matrix elements are given by
\begin{equation}
    \langle n, \mathbf{k}_\parallel \lvert \mathcal{H}_\mathrm{c} \rvert n', \mathbf{k}_\parallel\rangle = D_n\left(\mathbf{k}\right) \delta_{n, n'} - t_{n, n'} \left(\mathbf{k}_\parallel\right) \delta_{n, n' \pm 1}, \label{eq:MatrixElement}
\end{equation}
where
\begin{subequations}
    \label{eq:MatrixElementParts}
\begin{align}
	D_{i}\left(\mathbf{k}_\parallel\right) & = \frac{\hbar^2}{2 \Delta}\left[\frac{1}{m^+} + \frac{1}{m^-}\right] + V_i\left(\mathbf{k}_\parallel\right), \\
	t_{i, j}                               & = \frac{\hbar^2}{\left(m_i + m_j\right) \Delta^2}.
\end{align}
\end{subequations}
Note that here $m^+ = (m_i + m_{i +1}) / 2$ and $m^- = (m_i + m_{i-1}) / 2$, $m_i$ is the effective mass at mesh site $i$ and $\Delta$ is the (uniform) mesh-spacing.\\

\subsection{Self Energies for Coupling to Semi-Infinite Leads}
\label{appendixsec:Leads}
Coupling to external reservoirs is accounted for considering single-band electrons in a semi-infinite lead. At the device edge the discrete system represented by the closed system Hamiltonian needs to be truncated. The relevant component of the Schrodinger equation is of form
\begin{equation}
	E \eta_1 = - t_{0, 1} \eta_0 + D_1 \eta_1 - t_{1, 2} \eta_2,
\end{equation}
where $\eta_i$ is the wavefunction at lattice site $i$, $E$ is the energy and $t_{i, j}, D_i$ are the components of the Hamiltonian defined in Eq.~\ref{eq:MatrixElement}-~\ref{eq:MatrixElementParts}. To truncate the system to the region $i > 1$ it is necessary to eliminate $\eta_0$ from this equation. As the \emph{retarded} Green's function describes the response to an excitation generated within the device, the appropriate boundary condition is
\begin{equation}
	\eta_0 = \eta_1 e^{i k_1 \Delta},
\end{equation}
where $\Delta$ is the grid spacing at the device edge and $k_1$ is the electron wavevector at lattice site $1$. With this we can write
\begin{equation}
	E \eta_1 = - t_{0, 1} e^{i k_1 a} \eta_1 + D_1 \eta_1 - t_{1, 2} \eta_2.
\end{equation}
Note that in the contact $D$ encodes the boundary condition, and therefore the chemical potential in each lead.\\
A similar result can be found at the rightmost contact, leading to retarded self-energy
\begin{equation}
	\Sigma^R = \left(\begin{matrix}
		- t_{0, 1} e^{i k_1 \Delta} & 0      & \dots  & 0                        \\
		0                      & \ddots &        & \vdots                   \\
		\vdots                 &        & \ddots &                          \\
		0                      & 0      & \dots  & - t_{N, N+1} e^{i k_N \Delta}
	\end{matrix}\right).
\end{equation}
Finally the kinetic equation Eq.~\ref{eq:Kinetic} can be used to calculate the distributional Green's function $G^<$, which in turn gives the electron density through Eq.~\ref{eq:Charge}. Note that we achieve swift convergence in carrier density throughout the device by employing an adaptive Gauss-Kronrod integration scheme to discretize the energy space. Note that the distributional self-energies in the leads necessary to calculate the lesser and greater Green's functions can be calculated from the level broadening associated with the retarded Green's function, and the lead Fermi function as
\begin{equation}
    \Sigma^< = i f \Gamma.
\end{equation}

\subsection{Calculation of the Hartree potential}
\label{appendixsec:Potential}
Once the charge distribution $n$ has been calculated using Eq.~\ref{eq:Charge} the Hartree potential throughout the device can be found solving the Poisson equation subject to Neumann boundary conditions. For a system homogeneous in the $xy$-plane but inhomogeneous in $z$ this is given by
\begin{equation}
	\frac{\diff}{\diff z} \left[\epsilon\left(z\right) \frac{\diff \phi\left(z\right)}{\diff z}\right] + q \left[N_\mathrm{D}^+ \left(z\right) - N_\mathrm{A}^-\left(z\right) - n\left(z\right)\right] = F\left(\phi\right), \label{eq:Poisson}
\end{equation}
where $\epsilon$ is the static dielectric constant, $N_\mathrm{D}^- \; \left[N_\mathrm{A}^+\right]$ is the background donor (acceptor) density and $F$ is the residuum to be minimized. \\
Discretising on a nearest-neighbour tight-binding basis we can recover the non-zero matrix elements of the Poisson operator $\mathbf{A}$
\begin{subequations}
\begin{align}
	A_{ii} & = - \frac{\epsilon^+ + \epsilon^-}{\Delta^2} + q \left[N_{\mathrm{D}, i}^+ - N_{\mathrm{A}, i}^- - n_i\right], \\
	A_{ij} & = - \delta_{j, i\pm1} \frac{\epsilon_i + \epsilon_j}{2 \Delta^2},
\end{align}
\end{subequations}
where $\epsilon^\pm = (\epsilon_i + \epsilon_{i \pm 1})/ 2$, $\epsilon_i$ is the static dielectric constant at mesh point $i$ and $\Delta$ is the (uniform) mesh spacing. In matrix form we can write Eq.~\ref{eq:Poisson} as
\begin{equation}
	F\left[\phi\right] = \mathbf{A}\left[\phi\right] + q \left[N_\mathrm{D}^+ - N_\mathrm{A}^- - n\right].
\end{equation}
Using the current carrier density $n$ the quasi-Fermi level throughout the device is calculated by minimising
\begin{equation}
	n_i = N_\mathrm{C} \mathcal{F}_{1/2}\left[\frac{E_{\mathrm{F},i} - E_{\mathrm{C}, i} + q \phi_i}{k_\mathrm{B} T}\right].
\end{equation}
in which $T$ is the device temperature, $N_\mathrm{C}$ is the conduction band density of states, $\mathcal{F}_{m}$ is a Fermi-Dirac integral of order $m$, $E_{\mathrm{C}, i}$ is the conduction band offset at mesh site $i$, $E_{\mathrm{F}, i}$ is the quasi-Fermi level at mesh site $i$ and $n_i$ is the free charge calculated by the NEGF step. Changes in the free charge can then be \emph{predicted} for modest changes in potential $\delta \phi$
\begin{equation}
	n_{\mathrm{pred}, i} \left(\delta \phi_i\right) = N_\mathrm{C} \mathcal{F}_{1/2}\left[\frac{E_{\mathrm{F},i} - E_{\mathrm{C}, i} + q \left(\phi_i + \delta \phi_i\right)}{k_\mathrm{B} T}\right].
\end{equation}
At each iteration we solve for the \emph{change} in the electric potential $\delta \phi = \phi_\mathrm{out} - \phi_\mathrm{in}$ where $\phi_\mathrm{in}$ is the electric potential at the previous outer iteration
\begin{multline}
	F_\mathrm{pred} \left[ \delta \phi + \phi_\mathrm{in}\right] = \mathbf{A}\left[ \delta \phi + \phi_\mathrm{in}\right] \\
	+ q \left[N_\mathrm{D}^+ - N_\mathrm{A}^- - n_{\mathrm{pred}} \left(\delta \phi\right) \right] = 0.
\end{multline}
using a pre-conditioned Gauss-Newton predictor corrector scheme. The Jacobian can be calculated analytically as
\begin{equation}
	\mathbf{J}_{ij} = \frac{\diff F_{\mathrm{pred}, i}}{\diff \delta \phi_j} = A_{ij} + \frac{\diff n_{\mathrm{pred}, i}}{\diff \delta \phi_j},
\end{equation}
where the differential of the predictor charge density is given by
\begin{multline}
	\frac{\diff n_{\mathrm{pred}, i}}{\diff \delta \phi_j} = \delta_{i, j} \frac{q}{k_\mathrm{B} T} N_\mathrm{C} \\
	\times \mathcal{F}_{-1/2}\left[\frac{E_{\mathrm{F},i} - E_{\mathrm{C}, i} + q \left[\phi_i + \delta \phi_i\right]}{k_\mathrm{B} T}\right].
\end{multline}

\subsection{Self-Consistent Loop}
\label{appendixsec:Loop}
Convergence of an iterative fixed point calculation $x = f\left(x\right)$ is strongly dependent on how the input quantity is computed between iterations. A linear procedure, in which the previous input and output are combined as a simple mixture
\begin{equation}
	x^{k + 1} = \left(1 - \alpha\right) x^k + \alpha f\left(x^k\right),
\end{equation}
takes the weighting $\alpha$ as a model parameter. Such a method can be unstable, particularly in systems with narrow electronic resonances unless the weighting $\alpha$ is sufficiently small. Minimising the weighting too strongly however leads to unnecessarily slow convergence. This Paper employs an adaptive Type I Anderson mixing procedure to calculate subsequent inputs. In a generic Anderson mixing method inputs are calculated as linear mixtures of the outputs of previous iterations
\begin{align}
	x^{k + 1} = \sum_{j = 0}^{m_k} \alpha_j f\left(x^{k - m_k + j}\right),
\end{align}
in which $m_k$ is the memory capacity of the mixing scheme and $\alpha_k$ the mixing weights satisfying $\sum_{j = 0}^{m_k} \alpha_j = 1$. This can be recast as a minimisation
\begin{subequations}
\begin{align}
	 & \text{minimize}   & \lvert\lvert \sum_{j = 0}^{m_k} \alpha_j g\left(x^{k - m_k + j}\right) \rvert \rvert_2^2, \\
	 & \text{subject to} & \sum_{j = 0}^{m_k} \alpha_j = 1,
\end{align}
\end{subequations}
where $g\left(x\right) = x - f\left(x\right)$ is the residual of the fixed-point problem and the solution is the weight variable $\boldsymbol{\alpha}$. This Paper utilizes the stabilized Type I Anderson acceleration scheme outlined by Zhang \emph{et al.} \cite{zhangGloballyConvergentTypeI2020}.

\section{localized Phonon Self-Energies}
\label{appendix:B}
With a self-consistent potential $\phi$ it is possible to calculate the greater and lesser Green's functions at all energies and wavevectors. As discussed in Sec.~\ref{sec:GF} these Green's functions are related to correlation functions for electrons and unoccupied states in the conduction band. They can be utilized to calculate coupling between electronic states, mediated by an external reservoir. We can write the lesser self-energy for the reservoir to the lowest order perturbation as
\begin{align}
	\Sigma^\lessgtr\left(\mathbf{r, r'}; E\right) & = \int \mathcal{D}\left(\mathbf{r, r}'; \hbar \omega\right) G^\lessgtr\left(\mathbf{r, r}'; E \mp \hbar \omega\right) \mathrm{d} \hbar \omega, 
\end{align}
where $\mathcal{D}$ is a function describing the spatial correlation and energies of the scatterers
\begin{multline}
    \mathcal{D}\left(\mathbf{r, r'}; \hbar \omega\right) = \sum_\mathbf{Q} \lvert U_\mathbf{Q}\rvert^2 \\ \biggr[ e^{- i \mathbf{q} \cdot \left(\mathbf{r - r'}\right)} f_{q_z}^*\left(z\right) f_{q_z}\left(z'\right) N^0\left(\omega_\mathbf{Q}, T
		\right) \delta\left(\omega - \omega_\mathbf{Q}\right) \\
        + e^{i \mathbf{q} \cdot \left(\mathbf{r - r'}\right)} f_{q_z}(z) f_{q_z}^*(z') \left[N^0\left(\omega_\mathbf{Q}
	, T		\right) + 1\right] \delta\left(\omega + \omega_\mathbf{Q}\right)\biggr],
\end{multline}
in which $\mathbf{Q} = \left[\mathbf{q}, q_z\right]$ is the three-dimensional phonon wavevector, $U_\mathbf{Q}$ is the phonon potential felt by a single electron, $f_q$ is the out-of-plane mode profile and the Bose-Einstein distribution function as defined in Eq.~\ref{eq:BEDist}.

localized longitudinal optic phonon modes have a discrete out-of-plane wavevector space $\zeta_n = \frac{2 (n - 1) \pi}{d}, \; (n \in \mathbb{Z}^+$ where $d$ is the thickness of the nonlocal layer. Each branch $n$ follows dispersion Eq.~\ref{eq:localizedDispersion}. The quantized form of the localized phonon potential is given for the lowest order branch $n = 1$ by \cite{gubbinPolaritonicQuantizationNonlocal2022a}
\begin{equation}
	\hat{\phi}_\mathrm{L}\left(\mathbf{r}, z\right) = \frac{1}{e} \sum_{\mathbf{q}} U_{\mathbf{q}} u\left(z\right) e^{i \mathbf{q} \cdot \mathbf{r}} \left[\anlo{q} + \crlo{-q}\right].
\end{equation}
Here $\crlo{q} \; (\anlo{q})$ are creation (annihilation) operators for a phonon of transverse momentum $\mathbf{q}$ in the discrete mode and
\begin{align}
    U_{\mathbf{q}} & =\sqrt{\frac{e^2 \hbar \freqn{L}{q}}{\epsilon_{\mathbf{q}}^{\rho} \mathrm{A} d}} \frac{\sqrt{\lvert \mathbf{q} \rvert^2 + \zeta^2}}{\lvert \mathbf{q} \rvert^2 + \zeta^2 + q_0^2},
\end{align}
where $q_0$ is the screening length, $\mathrm{A}$ is the quantization surface and the dispersive Fr{\"o}hlich constant is given by
\begin{equation}
    \epsilon_{\mathbf{q}}^{\rho} = \epsilon_0 \epsilon_{\infty} \frac{\left(\freqn{L}{q}\right)^2}{\omega_{\mathrm{L}}^2 - \omega_{\mathrm{T}}^2},
\end{equation}
and the out-of-plane mode function is given by
\begin{equation}
	u\left(z\right) = \sin\left[\zeta \left(z - z_0 +d/2\right)\right].
	\label{eq:phonwave}
\end{equation}
in which $z_0$ defines the nonlocal layer center. The leading diagonal of the self-energies are given by
\begin{multline}
	\Sigma_{ii}^{\lessgtr} \left(\mathbf{k}, E\right) = \sum_{\mathbf{l}} \lvert U_{\mathbf{k-l}} \rvert^2 \lvert u(z_i)\rvert^2  \\
    \biggr\{ N^0\left(\freqn{L}{k-l}, T_a\right) G_{ii}^\lessgtr \left(\mathbf{l}, E \mp \hbar \freqn{L}{k-l} \right)\\
        \left[N\left(\freqn{L}{k-l}, T_a\right) + 1\right] G_{ii}^\lessgtr\left(\mathbf{l}, E \pm \hbar \freqn{L}{k-l} \right)
    \biggr\} \label{eq:diagse}
\end{multline}
where the subscripts indicate elements of the spatial Green's function or self-energy matrices and we changed wavevector to $\mathbf{l} = \mathbf{k - q}$. Using Eq.~\ref{eq:diagse} we can define the kernel for localized phonon scattering as
\begin{equation}
    \mathcal{K}_i^\mathrm{L}\left(\mathbf{q}\right) =  \lvert U_{\mathbf{q}} \rvert^2 \lvert u(z_i)\rvert^2. \label{eq:LOKernel}
\end{equation}
On passing to an integral over wavevector we recover
\begin{multline}
	\Sigma_{ii}^{\lessgtr} \left(\mathbf{k}, E\right) = \frac{e^2 \hbar}{2 \pi d}  \int l \diff \mathrm{l} \\
\frac{\freqn{L}{k-l}}{\epsilon_{\mathbf{k - l}}^{\rho}} \mathcal{G}\left(l, k, z_i\right) \biggr\{ N^0\left(\freqn{L}{k-l}, T_a\right) G^\lessgtr \left(\mathbf{l}, E \mp \hbar \freqn{L}{k - l} \right)\\
    \left[N^0\left(\freqn{L}{k-l}, T_a\right) + 1\right] G^\lessgtr\left(\mathbf{l}, E \pm \hbar \freqn{L}{k-l} \right)
	\biggr\}
\end{multline}
where we applied 
\begin{multline}
    \label{eq:angular_integral}
    \int_0^{2 \pi} \diff \theta \frac{k^2 + l^2 + 2 k l \cos \theta + \zeta^2}{\left[k^2 + l^2 + 2 k l \cos \theta + \zeta^2 + q_0^2\right]^2} \\
        =
	2 \pi \biggr[\frac{1}{\sqrt{\left(k^2 + l^2 + \zeta^2 + q_0^2\right)^2 - 4 k^2 l^2}}  \\
		- q_0^2 \frac{k^2 + l^2 + \zeta^2 + q_0^2}{\left[\left(k^2 + l^2 + \zeta^2 + q_0^2\right)^2 - 4 k^2 l^2\right]^{3/2}}\biggr],
\end{multline}
and defined the following
\begin{multline}
    \mathcal{G}\left(l, k, z\right) = u^2\left(z\right)   \\
	\times \biggr[\frac{1}{\sqrt{\left(k^2 + l^2 + \zeta^2 + q_0^2\right)^2 - 4 k^2 l^2}}  \\
		- q_0^2 \frac{k^2 + l^2 + \zeta^2 + q_0^2}{\left[\left(k^2 + l^2 + \zeta^2 + q_0^2 \right)^2 - 4 k^2 l^2\right]^{3/2}}\biggr].
\end{multline}

\bibliographystyle{naturemag} 
\bibliography{manuscript}

\end{document}